\documentclass[12pt]{article}
\usepackage{epsfig}
\usepackage{comment}
\usepackage{latexsym}
\usepackage{color}
\newcommand{\mysquare}[0]{\raise-.2ex\hbox{{\Large$\Box$}}}
\def\lsim{\mathrel{\rlap {\raise.5ex\hbox{$ < $}}
{\lower.5ex\hbox{$\sim$}}}}
\def\gsim{\mathrel{\rlap {\raise.5ex\hbox{$ > $}}
{\lower.5ex\hbox{$\sim$}}}} \topmargin -1.5cm \textheight=22.5cm \textwidth=16.5cm
\catcode`\@=11
\newcount\hour
\newcount\minute
\newtoks\amorpm
\hour=\time\divide\hour by60 \minute=\time{\multiply\hour by60 \global\advance\minute by-\hour}
\edef\standardtime{{\ifnum\hour<12 \global\amorpm={am}%
        \else\global\amorpm={pm}\advance\hour by-12 \fi
        \ifnum\hour=0 \hour=12 \fi
        \number\hour:\ifnum\minute<10 0\fi\number\minute\the\amorpm}}
\edef\militarytime{\number\hour:\ifnum\minute<10 0\fi\number\minute}
\def\draftlabel#1{{\@bsphack\if@filesw {\let\thepage\relax
   \xdef\@gtempa{\write\@auxout{\string
      \newlabel{#1}{{\@currentlabel}{\thepage}}}}}\@gtempa
   \if@nobreak \ifvmode\nobreak\fi\fi\fi\@esphack}
        \gdef\@eqnlabel{#1}}
\def\@eqnlabel{}
\def\@vacuum{}
\def\draftmarginnote#1{\marginpar{\raggedright\scriptsize\tt#1}}
\def\draft{\oddsidemargin -.2truein
        \def\@oddfoot{\sl preliminary draft \hfil
        \rm\thepage\hfil\sl\today\quad\militarytime}
        \let\@evenfoot\@oddfoot \overfullrule 3pt
        \let\label=\draftlabel
        \let\marginnote=\draftmarginnote
   \def\@eqnnum{(\theequation)\rlap{\k

 ern\marginparsep\tt\@eqnlabel}%
\global\let\@eqnlabel\@vacuum}  }

\newcommand{\be}[0]{\begin{equation}}
\newcommand{\ee}[0]{\end{equation}}
\newcommand{\ba}[0]{\begin{eqnarray}}
\newcommand{\ea}[0]{\end{eqnarray}}

\def\bs{\begin{subequations}}
\def\es{\end{subequations}}

\def\thebibliography#1{%
\vskip 0.5cm \centerline{\bf \Large References}
\list{%
[\arabic{enumi}]}{\settowidth\labelwidth{[#1]} \leftmargin\labelwidth \advance\leftmargin\labelsep
\usecounter{enumi}}
\def\newblock{\hskip .11em plus .33em minus .07em}
\sloppy\clubpenalty4000\widowpenalty4000 \sfcode`\.=1000\relax}

\renewcommand{\theequation}{\arabic{section}.\arabic{equation}}

\renewcommand{\section}{\setcounter{equation}{0}\@startsection
{section}{1}{0mm}{-\baselineskip}{0.5\baselineskip} {\normalfont\Large\bfseries}}

\renewcommand{\subsection}{\@startsection
{subsection}{2}{0mm}{-\baselineskip}{0.5\baselineskip} {\normalfont\large\bfseries}}

\renewcommand{\subsubsection}{\@startsection
{subsubsection}{3}{0mm}{-\baselineskip}{0.5\baselineskip} {\normalfont\normalsize\slshape}}

\usepackage{amsmath}
\usepackage{amssymb,amsfonts}
\usepackage{graphicx}
\usepackage{cite}

\renewcommand{\and}{\mbox{and}}

\begin{document}
\begin{titlepage}
\begin{flushright}
LPTENS--10/42,
October 2010
\end{flushright}


\begin{centering}
{\bf\huge String Vacua with Massive boson-fermion Degeneracy \& Non-Singular Cosmology}\\

\vspace{4mm}
 {\Large Ioannis~G.~Florakis \\
 }

\vspace{2mm}

$^1$ Laboratoire de Physique Th\'eorique,
Ecole Normale Sup\'erieure,$^\dagger$ \\
24 rue Lhomond, F--75231 Paris cedex 05, France\\
\vspace{2mm}
{\em  Ioannis.Florakis@lpt.ens.fr}

\vskip .1cm

 \vspace{3mm}

{\bf\Large Abstract}

\end{centering}
\begin{quote}

We discuss marginal deformations of string vacua with Massive boson-fermion Degeneracy Symmetry (MSDS), in connection to the cosmological evolution of the Universe from an early non-geometrical era. In particular, we discuss recent results on the stringy mechanism that resolves both Hagedorn divergences and the Initial Singularity problem.\\ 
\begin{center}
\emph{ Based on a talk given at the Workshop on Cosmology \& Strings, Corfu Institute, Greece, Sept 10, 2010}.
\end{center}

\noindent

\end{quote}
\vspace{5pt} \vfill \hrule width 6.7cm \vskip.1mm{\small \small \small \noindent $^\ast$
$^\dagger$\ Unit{\'e} mixte  du CNRS et de l'Ecole Normale Sup{\'e}rieure associ\'ee \`a
l'Universit\'e Pierre et Marie Curie (Paris
6), UMR 8549.}

\end{titlepage}
\newpage
\setcounter{footnote}{0}
\renewcommand{\thefootnote}{\arabic{footnote}}
 \setlength{\baselineskip}{.7cm} \setlength{\parskip}{.2cm}

\setcounter{section}{0}

\section{String Cosmology and Initial Vacuum Selection}

Some of the most challenging open problems of modern cosmology are directly related to the very early history of our Universe, where field theoretic notions are expected to break down due to the dominant quantum gravitational effects that arise in the hot and strongly curved phases of the early Universe. At the moment, string theory and its non-perturbative extension, M-theory, are the most promising candidates for a consistent theory of quantum gravity. To this end, a deep understanding of physics at the Planck scale calls for a purely stringy treatment \cite{CosmoTopologyChange}, which may help uncover the implications of this initial phase for the phenomenology and cosmology at lower energies and late times. 

Cosmological solutions arise naturally within the framework of perturbative string theory as quantum (or thermal) instabilities of an initially flat background. In particular, in string vacua with spontaneously broken supersymmetry, the non-vanishing genus-1 effective potential $V_{\textrm{eff}}(\mu_I)\neq 0$ will induce a backreaction on the underlying background. A correction of the tree-level background is then necessary in order to cancel the dilaton tadpole and restore conformal invariance at the 1-loop level. The same mechanism can be applied in the case of thermal vacua as well, where a finite free energy density  $\mathcal{F}(\beta)$ triggers a similar backreaction, with the various moduli $\mu_I$ and, in particular, the temperature $T=\beta^{-1}$, acquiring time dependence.

The finiteness of the 1-loop amplitude (free energy) allows the study of the emerging evolution within the framework of perturbative string theory and so, this mechanism lies at the core of string cosmology \cite{BV}. However, naive attempts to realize this attractive programme are generalically met with two major obstacles: (i) the Hagedorn divergences \cite{Hag}, corresponding to infinite backreaction and driving the theory through a non-perturbative regime and (ii) the Initial Singularity problem, traditionally plaguing standard cosmology. Furthermore, additional complications associated with the Vacuum selection mechanism may arise. Appart from the general demands for absence of Hagedorn/tachyonic divergences and gravitational singularities and for the perturbative tractability throughout the evolution, there are additional questions to be raised concerning the arbitrariness of the supersymmetry breaking mechanism. One may naturally ask the question of whether there exists a fundamental way to break supersymmetry; ideally, one that is dictated by symmetry principles. 

Of course, further constraints will need to be imposed on the Initial Vacuum, in order to ensure compatibility with late-time observational data. In particular, great care should be taken in the preparation of this initial state, so that it dynamically decompactifies at late times into a 3+1 Minkowski spacetime, with spontaneously broken $\mathcal{N}=1$ supersymmetric spectrum, 3 generations of chiral matter and a semi-realistic GUT gauge group, such as $SO(10)$.

A crucial first step in addressing these problems has been the discovery of a novel bose-fermi degeneracy symmetry in all massive modes of the theory, present at extended symmetry points in the moduli space of some special two-dimensional compactifications. This symmetry has been termed ``Massive Spectral bose-fermion Degeneracy Symmetry" (MSDS) \cite{MSDS}, \cite{reducedMSDS} and can be seen as a stringy enhancement of the current algebra of ordinary supersymmetry. In the simplest, maximally symmetric models, its degeneracy structure is manifested in terms of `generalized' Jacobi $\theta$-function identities, or in terms of the following identity of $SO(24)$ characters $V_{24}-S_{24}=24$. In particular its supersymmetric degeneracy structure at massive levels guarantees the absence of tachyonic excitations at the MSDS-points of moduli space.

In view of the above remarks, it is very natrual to consider the possibility that the very early universe arose as a hot compact ($d\leq 2$) space with curvature close to the string scale and that the underlying dynamics have driven some of the spatial dimensions to decompactify so that, eventually, a $d=4$ universe arises. Of course, the very early cosmological era is going to be characterized by a highly non-geometrical structure of the spacetime, demanding a treatment that properly takes into account the full, stringy degrees of freedom. In this respect, the high degree of symmetry of MSDS Vacua renders them natural candidates to describe this early stringy era. 

\section{Hagedorn Divergences and Non-Singular Cosmology}

As a first step in connecting $d\leq 2$ MSDS vacua with higher-dimensional supersymmetric vacua it is important to analyse their moduli space and study their tree-level adiabatic marginal deformations \cite{DeformedMSDS}. We consider here deforming the $\sigma$-model by operators of the current-current type $\int{d^2 z~\lambda_{\mu\nu}J^\mu(z) J^\nu(\bar z)}$. A careful examination of the deformed $\frac{SO(8,8)}{SO(8)\times SO(8)}$ compactification lattice in the maximally symmetric MSDS models indicates the presence of 2 independent moduli controlling the breaking of the left- and right- moving supersymmetries. These correspond to radial K\"{a}hler moduli associated to two specific toroidal cycles $X^0, X^1$. In the decompactification limit, the coupling to the $R$-symmetry charges is effectively washed out and one recovers a 4d, $\mathcal{N}=8$, type II vacuum and with an effective gauged supergravity description, the gauging being induced by the well-defined geometrical fluxes responsible for the breaking of supersymmetry.

The two additional large dimensions are seen to emerge from these deformations. In a cosmological setting, where the deformation moduli acquire time dependense, it becomes natural to consider our 4d cosmological space having been created dynamically from such an initial, 2d, MSDS vacuum. Of course, a necessary condition for this is for the initial vacuum to be free of tachyonic/Hagedorn divergences under arbitrary deformations of the dynamical moduli.	

Attempts to probe the early, high-temperature (and/or high-curvature) phase of the universe $R\sim\ell_s$, typically hit upon Hagedorn (/tachyonic) instabilities, preventing a perturbative treatment of the backreaction. Within the standard description of a thermal system in terms of a thermal trace, these divergences are seen to occur because of the exponential growth in the density of single-particle states with increasing masses. In the Euclidean picture, however, where time is compactified on a (toroidal) cycle $X^0$ of radius $R_0=\beta/2\pi$, the same phenomenon manifests itself slightly differently: namely, certain string states winding the Euclidean time cycle become tachyonic as soon as the thermal modulus $R_0$ exceeds a critical (Hagedorn) value $R_H$. From this point of view, Hagedorn divergences are not to be interpreted as true pathologies of string theory but, rather, as IR-instabilities of the Euclidean background and the corresponding phase transition is driven by tachyon condensation. The presence of these condensates injects non-trivial winding charge $\langle\mathcal{O}_n\rangle\neq 0$ into the vacuum, a property that will be crucial for our attempts to resolve these instabilities. Indeed, the correct dynamical treatment of the phase transition at the string level by condensing the thermal winding tachyon remains a formidable open problem. However, an alternative way recently proposed (see \cite{DeformedMSDS} and references therein), is to directly construct stable (non-tachyonic) thermal vacua with non-trivial winding charge, that would correspond to the resulting vacua describing the new phases after the transition. 

Actually, the thermal MSDS models correspond to temperatures higher than the Hagedorn value without running into divergences. This is possible because the Euclidean time cycle in these models is threaded by non-trivial ``gravito-magnetic" fluxes, associated to the $U(1)$ graviphoton and $U(1)$ axial vector gauge fields. Their presence refines the thermal ensemble and renders the free energy finite at the MSDS point. To illustrate this point, one may ``unfold" the fundamental domain (for sufficiently\footnote{This is required in order to ensure absolute convergence.} large $R_0^2\equiv G_{00}-G_{0I}G^{IJ}G_{J0}$) and decompose the integral into modular orbits. The $(0,0)$-orbit in winding space gives a vanishing contribution (because of supersymmetry at $T=0$) and the free energy is given by the following strip integral:
$$
	\int\limits_{||}{\frac{d^2\tau}{\tau_2^2}~(\,\ldots\,)~}\frac{R_0}{\sqrt{\tau_2}}\sum\limits_{\tilde{m}^0\neq 0}{e^{-\frac{\pi}{\tau_2}(R_0\tilde{m}^0)^2}(-)^{(a+\bar{a})\tilde{m}^0}}
		\sum\limits_{m_I,n^I\in\mathbb{Z}}{q^{\frac{1}{2}P_L^2}\bar{q}^{\frac{1}{2}P_R^2}(-)^{\bar{b}n^1}e^{2\pi i\tilde{m}^0(G_{0I}Q^I_{(M)}-B_{0I}Q^I_{(N)})}}~,
$$
where $Q^I$ are the 14 transverse $U(1)$ charges associated to the graviphoton and axial vector gauge fields. Let us now recall that the time component of a (constant) vacuum gauge potential $A_0$ cannot be gauged away in the presence of non-zero temperature. Its v.e.v. has physical meaning as a topological vacuum parameter that characterizes the thermal system.

In this sense, the thermal 1-loop amplitude (free energy) can be re-expressed \cite{DeformedMSDS} as a thermal trace over the Hilbert space of the initial 3d, $(4,0)$-theory:
$$
	Z(\beta,\mu^I,\tilde{\mu}_I)= \textrm{Tr}\left[~e^{-\beta H}~e^{2\pi i(\hat{G}_0^I \hat{m}_I-\hat{B}_{0I} n^I)}~\right],
$$
with $2\hat{m}_I, n^I$ being transverse integral charges. The thermal trace is deformed by the presence of the thermal fluxes associated to the graviphoton and axial vector. The combinations $\hat{G}_0^I\equiv G_{0K}G^{KI}$, $\hat{B}_{0I}\equiv B_{0I}-\hat{G}_0^K B_{KI}$ are scale-invariant, non-fluctuating, thermodynamical parameters of the thermal system. These (global) fluxes can be described as gauge field condensates, of zero field strength (locally), but with a non-vanishing value of the Wilson line around the Euclidean time cycle. At sufficiently low temperatures, charged states under these $U(1)$ fields acquire masses and effectively decouple from the thermal system, so that we effectively obtain the conventional thermal ensemble. Would-be tachyons can sometimes be lifted in this way, as they are charged under these gauge fields.

By a careful study of the low-lying thermal BPS-spectrum, it is possible to obtain a set of conditions \cite{DeformedMSDS} that guarantee the absence of Hagedorn divergences for any deformation of the transverse (dynamical) moduli. Whenever these are met, a discrete $O(8;\mathbb{Z})\times O(8;\mathbb{Z})$ transformation may rotate the conditions for the fluxes to the following convenient form:
$$
	\hat{G}_0^k=\hat{B}_{0k}=0~~,~~\hat{G}_0^1=2\hat{B}_{01}=\pm 1,
$$
with $k=2,\ldots, 7$ running over the toroidal directions transverse to the Euclidean time. In fact, these conditions have a very simple geometric meaning: they correspond to the case when the temperature cycle couples chirally to the left-moving $R$-symmetry charges and completely factorizes from the remaining toroidal manifold, which reamains coupled only to the right-moving fermion number $F_R$. Tachyon-free thermal-like models, thus, correspond to the following lattice factorization:
$$
	\Gamma_{(d,d)}[^{a\,,\,\bar{a}}_{b\,,\,\bar{b}}] = \Gamma_{(1,1)}[^a_b](R_0)\otimes \Gamma_{(d-1,d-1)}[^{\bar{a}}_{\bar{b}}](G_{IJ},B_{IJ}).
$$
Furthermore, in the thermodynamical phase saturating the above conditions, the deformed thermal trace simply reduces to the right-moving fermion index $	Z=\ln~\textrm{Tr}\left[~e^{-\beta H}(-)^{F_R}~\right]$.

It is easy to check that the maximally symmetric MSDS models fail to satisfy the above stability conditions, even though tachyon-free\footnote{Classes of tachyon-free MSDS models can also be obtained as asymmetric $\mathbb{Z}_2$-orbifolds of the maximally symmetric ones. In these, the fluctuations of dangerous moduli are projected out by the special structure of the orbifold action \cite{DeformedMSDS}.} trajectories connecting them to higher-dimensional $\mathcal{N}_4\leq 8$ vacua can still be identified for them. However, a very interesting class of tachyon-free thermal MSDS vacua are the thermal Hybrid models \cite{DeformedMSDS}, \cite{NonSingularCosmo}. In their cold $T=0$ version, these are $(4,0)$ vacua, with the right-moving supersymmetries spontaneously broken at the string scale and replaced by MSDS structure. This special (anti)chiral structure gives rise to an (enhanced) non-abelian gauge group $U(1)_L^8\times SU(2)^8_{k=2,R}$. Compactifying one of the longitudinal directions on a circle $S^1(R_0)$ and identifying it with the Euclidean time firection requires a special modding by the Scherk-Schwarz element $(-)^{F_L}\delta_0$, consistent with the spin-statistics connection. The free energy of the models then reads:
$$
Z=\frac{V_1}{8\pi}\int\limits_{\mathcal{F}}{\frac{d^2\tau}{\tau_2^{3/2}}}~\left[\frac{1}{2}\sum\limits_{a,b}{(-)^{a+b}\frac{\theta[^a_b]^4}{\eta^4}}\right]~\Gamma_{E_8}~(\bar{V}_{24}-\bar{S}_{24})\Gamma_{(1,1)}[^a_b](R_0).
$$
where $\Gamma_{(1,1)}[^a_b](R_0)= \frac{R_0}{\sqrt{\tau_2}}\sum\limits_{\tilde{m}^0,n^0}{e^{-\frac{\pi R_0^2}{\tau_2}|\tilde{m}^0+\tau n^0|^2}(-)^{\tilde{m}^0 n^0+a\tilde{m}^0+b n^0}}$. The conditions for absence of tachyon divergences are saturated by the factorized structure of the $\Gamma_{(1,1)}$ lattice and the free energy is finite for any value of the transverse dynamical moduli. Furthermore, the factorization is preserved, since the mixing moduli (associated to the gravitomagnetic fluxes) do not correspond to fluctuating fields. In addition, the presence of these particular fluxes restores the thermal $T$-duality\footnote{ The duality corresponds to $R_0\rightarrow 1/(2R_0)$, together with a simultaneous interchange of the spinor chiralities $S_8\leftrightarrow C_8$.} and injects non-trivial winding charge into the thermal vacuum.

It is important to note that the Euclidean path integral reduces to the expression for the deformed thermal trace $\textrm{Tr}{ e^{-\beta H}~(-)^{F_R}}$ only for the region $R_0>1/\sqrt{2}$, where absolute convergence is ensured. In order to obtain the analogous expression for $R_0<1/\sqrt{2}$ one should first perform a double Poisson resummation to go to the dual phase and then unfold the fundamental domain. Thus, the thermal trace description is an example of a field theoretic expression that fails to capture the thermal duality symmetry. This hints again to the fact that the fundamental object is the Euclidean path integral, which is valid for all temperatures and which exhibits the stringy dualities manifestly.

An important result \cite{DeformedMSDS} is that, whenever the MSDS structure is preserved in the thermal Hybrid models, MSDS symmetry together with level matching permit the explicit calculation of the thermal 1-loop amplitude:
$$
	\frac{Z}{V_1} = 24\times\left(R_0+\frac{1}{2R_0}\right)-24\times\left|R_0-\frac{1}{2R_0}\right|.
$$
This result is exact at the genus-1 level, without any $\alpha'$-approximation. It manifestly exhibits the celebrated thermal duality and the non-analytic (conical) structure in the second term is induced by the presence of extra massless states at the self-dual fermionic point $R=1/\sqrt{2}$. Futhermore, preservation of MSDS structure introduces well-defined cancelations between the massive towers of states of the cold theory, so that the (deformed) thermal trace can be seen to reduce to a conventional thermal ensemble $\left.\textrm{Tr}\right|_{m^2=0}{e^{-\beta H}}$, restrained over the massless Hilbert space of the cold $(4,0)$ theory. In view of this result, it comes as no surprise that the thermal equation of state defined by these models $\rho=P=48\pi T^2$ is exactly that of thermal massless radiation in 2d \cite{DeformedMSDS}, \cite{NonSingularCosmo}.

This observation implies the expression $T=T_c e^{-|\sigma|}$ of the physical, duality-invariant temperature in terms of a thermal variable $\sigma\in(-\infty,+\infty)$, where now $T_c \equiv \frac{\sqrt{2}}{2\pi}$ is the critical maximal temperature for the theory. As the $\sigma$-parameter increases, one might envisage the system heating up, reaching the critical temperature, at which point it undergoes a phase transition and, subsequently, cooling down in the new phase. The presence of the phase transition can be motivated as follows \cite{NonSingularCosmo}. Consider a pure momentum state in the $S_8 \bar{V}_{24}$ sector at the extended symmetry (critical) point $\sigma=0$. At this point, one observes the presence of localized operators inducing transitions between pure momentum and pure winding states. For example, taking the zero mode of the current $J_{-}=\psi^0 e^{-iX_L^0}$ and acting on the vertex operator of the pure momentum state:
$$
	J_{-}(z)e^{-\phi/2}S_{10,\alpha}e^{\frac{i}{2}X_L^0+\frac{i}{2}X_R^0}\bar{V}_{24}(w)\sim \frac{1}{z-w}\gamma^0_{\alpha\dot\beta}e^{-\phi/2}C_{10,\dot\beta}e^{-\frac{i}{2}X_L^0+\frac{i}{2}X_R^0}\bar{V}_{24}(w),
$$
one obtains a pure winding state of opposite chirality, in the $C_8\bar{V}_{24}$ sector. Therefore, the extended symmetry point $\sigma=0$ is characterized by the presence of non-trivial 3-point amplitudes conducting transitions between pure momentum and pure winding states.

This phase transition may be shown to admit an effective description in terms of a spacelike brane \cite{NonSingularCosmo}, gluing together the space of momenta with the dual space of windings. The effective action,
$$
S=\int{d^2 x~e^{-2\phi}\sqrt{-g}\left(\frac{1}{2}R+2(\nabla\phi)^2\right)}+\int{d^2 x~\sqrt{-g}~P}-\kappa\int{dx^1 d\sigma~e^{-2\phi}\sqrt{g_{11}}\delta(\sigma)},
$$
in addition to the gravity-dilaton and the thermal effective potential terms, contains a spacelike brane contribution localized at the phase transition $\sigma=0$. It can be seen as representing the localized negative pressure, sourced by the 24 extra (complex) massless scalars at the extended symmetry point. Imposing conservation of the thermal entropy across the transition (absence of latent heat) leads to the following cosmological solution \cite{NonSingularCosmo} in the conformal gauge:
$$
ds^2 = \frac{4}{\kappa^2}~\frac{e^{|\tau|}}{1+|\tau|}~\left(-d\tau^2+dx^2~\right)~~~~,~~~~g_{s}^2 \equiv e^{2\phi(\tau)}=\frac{\pi\kappa^2}{192}~\frac{1}{1+|\tau|}.
$$
Thus, it can be seen that the presence of the phase transition at the extended symmetry point induces a bounce both in the scale factor and the dilaton and the cosmological evolution evades the gravitational singularity, while remaining within the perturbative regime, provided that $\kappa^2\ll 1$. This is the first example in the literature where, a careful treatment of the stringy degrees of freedom around the extended symmetry point, uncovers a stringy mechanism that simultaneously resolves both the initial singularity as well as the Hagedorn divergences. It is expected that the basic ingredients of this mechanism might protect the evolution from singularities in higher dimensional vacua as well. Obviously, this is only the first step in the ambitious programme to connect the early non-singular era of the Universe with the standard late-time cosmology and phenomenology.

\section{Acknowledgements}

	I would like to thank C.~Kounnas, H.~Partouche and N.~Toumbas for sharing several fruitful discussions during our collaboration. Furthermore, it is a pleasure to thank the organizers for giving me the opportunity to speak at the Corfu Workshop on Cosmology and Strings 2010.
	


\begin{thebibliography}{10}


\bibitem{CosmoTopologyChange}  
  E.~Kiritsis and C.~Kounnas,
  ``Dynamical topology change, compactification and waves in string
  cosmology,''
  Nucl.\ Phys.\ Proc.\ Suppl.\  {\bf 41} (1995) 311.

\bibitem{BV}
  R.~H.~Brandenberger and C.~Vafa,
  ``Superstrings in the Early Universe,''
  Nucl.\ Phys.\  B {\bf 316}, 391 (1989).

\bibitem{Hag}
	R.~Hagedorn, ``Statistical thermodynamics of strong interactions at high-energies," 
	Nuovo Cim. Suppl. {\bf 3}, 147 (1965).
	
\bibitem{MSDS}
  C.~Kounnas,
  ``Massive boson-fermion degeneracy and the early structure of the universe,'' 
  Fortsch.\ Phys.\ {\bf 56} 1143 (2008).

\bibitem{reducedMSDS}
  I.~Florakis and C.~Kounnas,
  ``Orbifold Symmetry Reductions of Massive Boson-Fermion Degeneracy,'' 
  Nucl.\ Phys.\ B {\bf 820} 237 (2009).
  
\bibitem{DeformedMSDS}
	I.~Florakis, C.~Kounnas and N.~Toumbas,
	``Deformations of Vacua with Massive boson-fermion Degeneracy Symmetry,''
	Nucl.\ Phys.\ B {\bf 834} 273 (2010).

\bibitem{NonSingularCosmo}
	I.~Florakis, C.~Kounnas, H.~Partouche and N.~Toumbas,
	``Non-singular string cosmology in a 2d Hybrid model,''
	Preprint CPHT-RR077.0810, LPTENS-10-31, (2010).
	
\end{thebibliography}
\end{document}